\documentclass[article]{jss}

\makeatletter
\def\@myoddhead{%
  {\color{white} JSS}\\[-1.42cm]
  \hspace{-2em}\includegraphics[height=23mm,keepaspectratio]{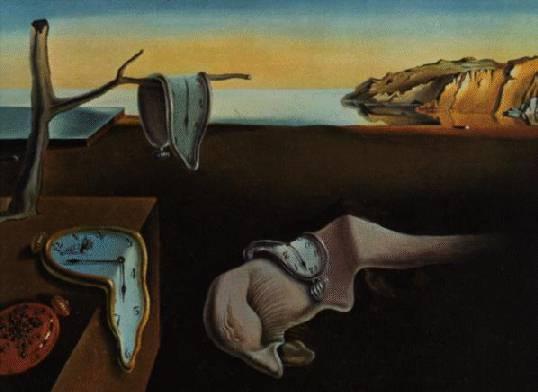}\hfill
  \parbox[b][23mm]{118mm}{\hrule height 3pt
    \centering\vspace{4mm}
    {\fontfamily{pzc}\fontsize{28}{32}\selectfont Journal of Statistical Software}
    \par\vspace{4mm}\hrule height 3pt}}
\renewcommand{\makefooter}{%
  \vspace{\footerskip}
  \begin{samepage}
  \textbf{\large \@Footername:}\\[.3\baselineskip]
  \@Address
  \end{samepage}}
\makeatother
\usepackage{amsmath,amssymb,booktabs,needspace,xurl,array,float}
\newcolumntype{P}[1]{>{\raggedright\arraybackslash}p{#1}}
\AddToHook{cmd/makefooter/before}{\Needspace{24\baselineskip}}
\graphicspath{{figures/}}
\newcommand{\maintainername}{Valerie Le}
\newcommand{\contactname}{Valerie}
\newcommand{\FirstBoundaryP}{0.00191}
\newcommand{\SecondBoundaryP}{0.00789}

\newcommand{\SubmittedHistoryRows}{%
1 & $H_1$ & 205 & 2.8922 & 0.00191 & 0.090 & Do not reject \\
2 & $H_1$ & 285 & 2.4141 & 0.00789 & 0.006 & Reject \\
3 & $H_2$ & 393 & 1.9600 & 0.02500 & 0.020 & Reject \\
}

\newcommand{\RecipientRows}{%
1 & 1 & $H_1$ & 0.0125 & 33 & 0.5000 & 0.0000137 & Retain \\
2 & 1 & $H_2$ & 0.0125 & 33 & 0.5000 & 0.0000137 & Retain \\
3 & 2 & $H_1$ & 0.0125 & 67 & 0.0080 & 0.0022725 & Retain \\
4 & 2 & $H_2$ & 0.0125 & 67 & 0.0001 & 0.0022725 & Reject \\
5 & 2 & $H_1$ & 0.0250 & 67 & 0.0080 & 0.0061421 & Retain \\
6 & 3 & $H_1$ & 0.0250 & 100 & 0.0200 & 0.0230919 & Reject \\
}

\newcommand{\RecipientOldCutoff}{0.0022725}
\newcommand{\RecipientNewCutoff}{0.0061421}

\author{Valerie Le\\Rutgers University\\New Brunswick, New Jersey
  \And MengYang Yi\\Johns Hopkins University
  \AND Philip He\\Celcuity}
\Plainauthor{Valerie Le, MengYang Yi, Philip He}
\title{\pkg{graphMTP}: An \proglang{R} Shiny Interface for Graphical Multiple Testing and Interim Monitoring}
\Plaintitle{graphMTP: An R Shiny Interface for Graphical Multiple Testing and Interim Monitoring}
\Shorttitle{\pkg{graphMTP}: Graphical Multiple Testing}
\Abstract{
Graphical multiple testing procedures provide a framework for controlling the
family-wise error rate across endpoints, treatment comparisons, and patient
populations in confirmatory clinical trials. Interim monitoring requires
analysts to track changing local significance levels together with
hypothesis-specific testing schedules. We present \pkg{graphMTP}, an
\proglang{R} application built with \pkg{shiny}, for
constructing hypothesis graphs, specifying per-hypothesis group-sequential
plans, and inspecting alpha recycling. The application uses
\pkg{TrialSimulator} for graph updates, \pkg{gsDesign} for spending
profiles, and \pkg{rpact} for boundary calculations. It records the local levels, boundaries, and decisions used for
each submitted test and exports designs and histories as JSON files.
Worked examples cover fixed-design reallocation, fixed-sequence gatekeeping,
delayed activation, and retesting a previously monitored recipient of alpha.
Replication code checks graph updates, normal crossing probabilities, and
saved-session replay. Sequential tests use a prespecified information schedule and reconstruct
the complete local boundary family at the current allocation, including
after recycling. Historical event records retain the boundaries actually used.
An installable package and standalone sources accompany the article.}
\Plainkeywords{graphical multiple testing, family-wise error rate, group-sequential design, clinical trials, R, shiny}
\Address{Valerie Le\\Rutgers University\\New Brunswick, New Jersey\\
E-mail: \email{hpl14@scarletmail.rutgers.edu}\\[8pt]
MengYang Yi\\Johns Hopkins University\\[8pt]
Philip He (corresponding author)\\Celcuity\\
E-mail: \email{philipyihe2018@gmail.com}\\[8pt]
Software maintainers: Valerie Le and MengYang Yi.\\
Package contact: Valerie Le.}
\begin{document}
\raggedbottom
% Avoid the legacy class's over-wide centreline marks on recent LaTeX kernels.
\markboth{\pkg{graphMTP}: Graphical Multiple Testing}{Valerie Le, MengYang Yi, Philip He}
\section{Introduction}

In confirmatory clinical trials, investigators frequently evaluate multiple
hypotheses simultaneously, such as hypotheses concerning several endpoints,
treatment comparisons, or patient populations. Testing each at an unadjusted
significance level can increase the probability of at least one false positive
conclusion. Multiplicity procedures address this problem by controlling the
family-wise error rate (FWER), the probability of rejecting at least one true
null hypothesis in the specified family \citep{FDA2022,EMA2002}.

Classical methods include Bonferroni testing, Holm's step-down procedure
\citep{Holm1979}, and fixed-sequence testing. Clinical objectives can also
require fallback or gatekeeping strategies, in which the testing of one
endpoint depends on results for another \citep{Wiens2003,Dmitrienko2003}.
The ordering encodes priorities among clinical claims.

The graphical approach represents a multiple-testing strategy by nodes with
initial local significance levels and directed edges with transition weights
\citep{Bretz2009}. Rejection removes a node from the active testing graph and redistributes its level
according to graph-updating rules. Within the Bonferroni-based framework, the
resulting sequentially rejective procedure corresponds to a consonant closed
test and provides strong FWER control under its assumptions
\citep{Marcus1976,Bretz2011}. Group-sequential extensions combine graphical
weights with repeated local tests and error spending across interim analyses
\citep{LanDeMets1983,MaurerBretz2013}.

The methodological framework is established, but specifying a graph and
checking the consequences of successive rejections still require care.
Interim monitoring adds a separate schedule for each hypothesis, information
fractions, and spending rules. Applied statisticians must check the graph
updates and monitoring plan and explain the resulting strategy to collaborators who may
not work directly from transition matrices.

Several packages address these tasks. \pkg{gMCP} provides graphical multiple
testing and a desktop editor \citep{gMCP2024}; \pkg{gMCPShiny} provides a
browser interface and hosted applications \citep{gMCPShiny2023}.
\pkg{graphicalMCP} implements graphical procedures programmatically and,
in version 0.3.0, also documents group-sequential testing
\citep{graphicalMCP2026}. \pkg{gsDesign} and \pkg{rpact} provide
group-sequential design calculations \citep{gsDesign2026,rpact2026}.
An interactive workflow can connect these design calculations to the
successive inputs and decisions that a clinical team needs to review.

We present \pkg{graphMTP}, a \pkg{shiny} application that connects interactive
graph construction to a per-hypothesis monitoring plan and a saved record of
submitted analyses. Users specify local alpha allocations and transition
weights, inspect planned boundaries, and submit one-sided $p$~values and
observed information counts. The application records each testing event and
supports JSON export and replay. We describe its implementation, illustrate
four worked cases, and provide replication code and verification cases.
The examples document graph updates, sequential boundaries, and saved
analysis histories produced by this workflow.

\Needspace{10\baselineskip}
\section{Statistical background}
\label{sec:background}

\subsection{Family-wise error rate and closed testing}

Consider the problem of testing $m$ elementary null hypotheses
$H_1, \ldots, H_m$, and let
\[
I = \{1, \ldots, m\}
\]
denote the associated index set. In confirmatory clinical trials, the
family-wise error rate (FWER), defined as the probability of rejecting at least
one true null hypothesis, is typically required to be controlled in the strong
sense at a prespecified level $\alpha$: control holds under every
configuration of true and false elementary null hypotheses.

A general principle for constructing such procedures is the closed testing
principle of \citet{Marcus1976}. For every nonempty subset $J \subseteq I$,
define the intersection hypothesis
\[
H_J = \bigcap_{j \in J} H_j.
\]
The corresponding closed testing procedure rejects an elementary hypothesis
$H_i$, $i \in I$, if and only if all intersection hypotheses $H_J$ with
$i \in J \subseteq I$ are rejected by their corresponding local tests,
each having size at most $\alpha$ under its intersection null.
By construction, the resulting procedure controls the FWER in the strong sense
at level $\alpha$. However, direct implementation becomes cumbersome as the
number of hypotheses increases, because the closure contains $2^m - 1$
nonempty intersection hypotheses.

\subsection{Graphical multiple testing procedures}

An important class of closed testing procedures is obtained by applying
weighted Bonferroni tests to the intersection hypotheses
\citep{Bretz2009,Bretz2011}. For each nonempty subset $J \subseteq I$, assume
a collection of weights $w_j(J)$, $j \in J$, satisfying
\[
0 \leq w_j(J) \leq 1
\qquad \text{and} \qquad
\sum_{j \in J} w_j(J) \leq 1.
\]
The weighted Bonferroni test rejects $H_J$ if
\[
p_j \leq w_j(J)\alpha
\]
for at least one $j \in J$, where $p_j$ denotes the unadjusted $p$~value for
$H_j$, assumed valid under that elementary null.

Graphical multiple testing procedures provide a convenient representation of
such weighting strategies. Following \citet{Bretz2011}, a graphical weighting
strategy is specified through the weights $w_i(I)$, $i \in I$, for the global
intersection hypothesis $H_I$, together with a transition matrix
\[
G = (g_{ij}),
\]
where
\[
0 \leq g_{ij} \leq 1, \qquad g_{ii} = 0,
\qquad \text{and} \qquad \sum_{j=1}^m g_{ij} \leq 1
\]
for all $i,j \in I$. Here, $w_i(I)$ denotes the initial weight assigned to
$H_i$, and $g_{ij}$ indicates the fraction of the local significance level at
$H_i$ that is propagated to $H_j$ if $H_i$ is rejected.

The full collection of weights $w_j(J)$ for the closure is induced from
$w_i(I)$ and $G$ through the graph-updating algorithm described by
\citet{Bretz2011}. Many standard multiple testing procedures, including
weighted Bonferroni, Holm, fixed-sequence, fallback, and gatekeeping
procedures, can be expressed within this framework
\citep{Bretz2009,Bretz2011}.

For Bonferroni-based graphical procedures, consonance and sequential rejection
play an important role. In particular, if the monotonicity condition
\[
w_j(J) \leq w_j(J')
\qquad \text{for all } J' \subseteq J \subseteq I
\text{ and } j \in J'
\]
is satisfied, then the resulting closed testing procedure is consonant
\citep{Hommel2007,Bretz2011}. This leads to a sequentially rejective shortcut
procedure, so that the graphical multiple test can be carried out through
iterative graph updates rather than explicit evaluation of all
$2^m - 1$ intersection hypotheses.

For a rejection of $H_r$, the updated local levels of surviving hypotheses
are
\begin{equation}
\alpha_j'=\alpha_j+\alpha_r g_{rj}, \qquad j\ne r.
\label{eq:alpha-update}
\end{equation}
For distinct surviving $i,j$, the transition update is
\begin{equation}
g_{ij}'=
\begin{cases}
\dfrac{g_{ij}+g_{ir}g_{rj}}{1-g_{ir}g_{ri}},&g_{ir}g_{ri}<1,\\[5pt]
0,&g_{ir}g_{ri}=1,
\end{cases}
\qquad g_{ii}'=0.
\label{eq:edge-update}
\end{equation}
These updates use the graph immediately before the rejection
\citep{Bretz2009}. The total level remaining in the graph cannot exceed
the total level before rejection.
It is conserved when the rejected node redistributes its full level; a
terminal node with no outgoing weight can remove level from the graph.
The initial absolute levels satisfy $\alpha_i^{(0)}=\alpha w_i(I)$;
subsequent levels correspond to weights on the surviving set.
The application stores absolute local levels, whereas $w_i(J)$ and
$g_{ij}$ are dimensionless.

\subsection{Group-sequential extension}
\label{sec:sequential-scope}

Let hypothesis $H_i$ have planned information fractions
$0<t_{i1}<\cdots<t_{iK_i}=1$, where $K_i$ is its number of looks.
A look is an analysis of one hypothesis at a specified data cut.
For a fixed local level $\gamma$, let $a_i(\gamma,t)$ denote cumulative
error spending, with $a_i(\gamma,0)=0$, $a_i(\gamma,1)=\gamma$, and
nondecreasing spending in $t$. Under the canonical joint-normal model,
the repeated statistics for $H_i$ have zero means and unit variances at
the null boundary, with
\begin{equation}
\operatorname{Corr}(Z_{is},Z_{it})
 =\sqrt{t_{is}/t_{it}},\qquad s\leq t.
\label{eq:canonical-correlation}
\end{equation}
The null boundary is assumed to be least favorable within each elementary
null. For one-sided composite nulls in the joint-normal model, a nonpositive
mean vector and covariance constant over the null make upper-boundary
crossing no more likely than at mean zero.
For endpoint-specific tests, this may be an asymptotic approximation.
Valid marginal $p$~values alone do not establish this joint distribution.

Writing $p_{is}=1-\Phi(Z_{is})$, the nominal cutoffs
$\alpha^*_{is}(\gamma)$ are chosen so that
\begin{equation}
P_{0,i}\!\left(p_{is}\leq\alpha^*_{is}(\gamma),\
p_{iu}>\alpha^*_{iu}(\gamma)\text{ for all }u<s\right)
=a_i(\gamma,t_{is})-a_i(\gamma,t_{i,s-1}),
\label{eq:incremental-spending}
\end{equation}
where $t_{i0}=0$ and $P_{0,i}$ is probability at the null boundary.
The spending increment is the probability of first crossing at that look;
the nominal cutoff is the threshold applied to the observed $p$~value
\citep{LanDeMets1983}.

\citet{MaurerBretz2013} extend Bonferroni-based graphical testing by
constructing sequential local tests at the intersection-specific levels
$\alpha w_i(J)$. The sequentially rejective shortcut requires graph-weight
monotonicity and nested local rejection regions. At fixed information times,
a sufficient boundary condition is
\begin{equation}
\alpha^*_{ik}(\gamma')\geq\alpha^*_{ik}(\gamma),
\qquad \gamma'\geq\gamma,\quad k=1,\ldots,K_i.
\label{eq:boundary-nesting}
\end{equation}
One sufficient spending condition is that every increment
$a_i(\gamma,t_{ik})-a_i(\gamma,t_{i,k-1})$ is nondecreasing in $\gamma$.
The first-crossing recursion then gives nested nominal cutoffs by induction.
Fixed cumulative proportions $a_i(\gamma,t_{ik})=\gamma c_{ik}$,
$0<c_{i1}<\cdots<c_{iK_i}=1$, satisfy this condition. This includes
Lan--DeMets Pocock and fixed-shape Hwang--Shih--DeCani profiles.
Lan--DeMets O'Brien--Fleming spending also satisfies the increment condition
over $0<\gamma\leq2\{1-\Phi(1)\}\approx0.3173105$, which contains
the supported local levels.
These conditions concern a prespecified family of local tests evaluated
at each allocation. Bonferroni protection permits dependence between
hypotheses; Equation~\ref{eq:canonical-correlation} specifies the separate
within-hypothesis requirement.

To see the FWER implication, let $T$ be the set of true null hypotheses.
Before the first true rejection, the surviving set $J$ contains $T$, and
$\alpha w_i(J)\leq\alpha w_i(T)$ for $i\in T$. Nesting places any true
rejection inside the union of the local rejection regions for the true
hypotheses at levels $\alpha w_i(T)$. The union bound then gives probability
at most $\alpha\sum_{i\in T}w_i(T)\leq\alpha$. This argument permits
arbitrary dependence between hypotheses and requires the stated local
joint model within each hypothesis. When that model is asymptotic, the
corresponding error-control statement is asymptotic.

The hypothesis family, graph, complete information schedule, and spending
profiles are prespecified. At every look, including the final look, the
application reconstructs the complete local boundary family at the current
allocation (Section~\ref{sec:impl-integration}). Submitted information must
match the locked schedule. The application compares only the current
$p$~value with its reconstructed cutoff; it does not retrospectively reject
on an earlier $p$~value. This restriction can omit rejections available to a
procedure that looks back over all observed looks. The resulting decisions
must therefore be distinguished from the full closed-testing procedure.
The supported scope excludes changes to the information schedule, futility
stopping, and outcome-dependent redesign.

\section[Overview of graphMTP]{Overview of \pkg{graphMTP}}
\label{sec:software}

\subsection{Design goals and scope}
\label{sec:goals}

\pkg{graphMTP} is an interactive application built with \pkg{shiny}
for applied statisticians and clinical-trial teams who
need to specify and inspect a graphical testing strategy. The application
favours direct manipulation: hypotheses are drawn as nodes, alpha is allocated
on those nodes, and the effect of each rejection is shown by updated levels
and directed edges. A design and its submitted analysis history can be saved
in a JSON file for review or reuse \citep{shiny2026}.

The application takes a prespecified hypothesis family, graph, monitoring
plan, and externally calculated one-sided $p$~values. Information counts
must be proportional to statistical information for the canonical
joint-normal approximation in Section~\ref{sec:background}; a patient count
requires an endpoint-specific justification. Whole-number counts are an
interface and backend restriction, rather than a mathematical requirement
of information-based sequential testing.

\subsection{Main workflow}
\label{sec:workflow}

The four tabs follow the stages of constructing and executing a procedure.
The \code{Home} tab provides an overview and references. On the
\code{Design} tab, right-clicking the canvas adds a node, double-clicking a
node edits its label and local significance level, and directed edges specify
how alpha is redistributed. Users specify the intended family level through
their allocation: $\sum_i\alpha_i^{(0)}\leq\alpha$.
For a one-sided family level of $0.025$, they allocate at most $0.025$.
The interface checks only that the total is at most one.

\code{Create Object} constructs the testing object. In the fixed-design
workflow, \code{Reject Selected} applies a user-selected rejection and shows
the resulting graph update. This control explores the consequences of a
rejection; it does not evaluate an observed $p$~value. Users must establish
that the hypothesis is rejectable at its current local level before using
that update as part of an actual analysis.

\begin{figure}[H]
\centering
\includegraphics[width=\linewidth]{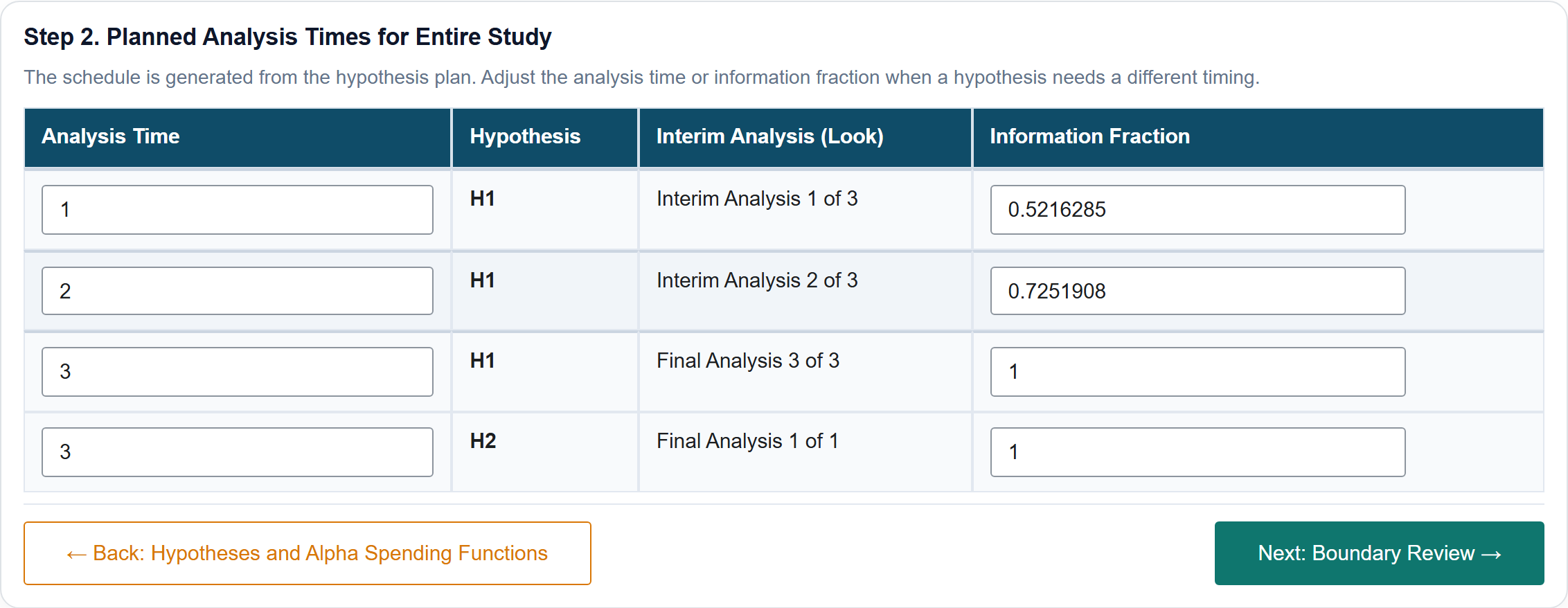}
\caption{\label{fig:app-overview} Monitoring-schedule view of \pkg{graphMTP}.
Each hypothesis look has an analysis-time index and planned information
fraction. The primary's three looks and the secondary's single final look
define the example in Section~\ref{sec:ex-gs}.}
\end{figure}

The \code{Group Sequential Design} tab specifies each hypothesis's planned
number of looks, maximum information count, and spending rule. Its next step
maps each look to a shared analysis time and a planned information fraction.
The shared analysis time is an integer data-cut or submission-batch index,
not elapsed calendar time (Figure~\ref{fig:app-overview}).
The final step displays boundary previews before \code{Finalize Design}.
The available choices are Lan--DeMets O'Brien--Fleming, Lan--DeMets Pocock,
Hwang--Shih--DeCani, an HP-derived profile, and custom cumulative spending.
Section~\ref{sec:impl-integration} gives their exact runtime parameterisations.

On the \code{Analysis} tab, users submit one-sided $p$~values and positive
whole-number observed information counts for the active hypotheses at the
earliest available analysis time. Information must match the locked planned count and increase within each
hypothesis. Alpha
recycling can trigger additional testing within that analysis time.
All currently active scheduled hypotheses are submitted together. A hypothesis
with zero alpha is excluded from the input batch; if it becomes active,
its remaining scheduled inputs can be submitted in a subsequent batch at
that analysis time. Recycling within a batch can trigger a retest using
the same $p$~value and information, without creating a new data look. The
\code{Submitted Analyses} table records the boundary, local alpha,
$p$~value, and decision for each testing event, including retests.

Previews and submitted tests use the same boundary construction and effective
information fractions derived from the planned counts. Recycling can change
the allocation between a preview and a submitted test. A testing event is a
single recorded boundary comparison, so one hypothesis look can produce
several event rows after recycling.

\Needspace{8\baselineskip}
\section{Software implementation}
\label{sec:implementation}

\subsection{Application architecture}
\label{sec:impl-arch}

The interface is organised as a \pkg{shiny} \code{navbarPage}.
The tab definitions and server logic are separated into sourced files for
graph editing, schedule validation, boundary calculation, analysis submission,
and display. The source components share one server environment. \pkg{visNetwork}
\citep{visNetwork2025} renders the graph, \pkg{DT} displays tables,
and \pkg{shinyjs} supports client-side interactions.

A \code{reactiveValues} store holds the session state. Two tibbles contain
the editable nodes and edges, while separate tables contain the hypothesis
plans, scheduled looks, and submitted history. The
\code{TrialSimulator::GraphicalTesting} object maintains the live testing
state. Keeping editable design tables distinct from the testing object allows
the application to validate a design before execution and to preserve the
inputs needed for replay.

The installable \pkg{graphMTP} package provides the launcher
\code{run\_app()}. The package build copies the maintained application sources
into its installed application directory, so the standalone scripts and
package distribution use the same implementation. The existing development
entry point remains \code{app.R}. The launcher can be called from any working
directory:

\begin{CodeChunk}
\begin{CodeInput}
R> library("graphMTP")
R> run_app()
\end{CodeInput}
\end{CodeChunk}

\subsection{Graph representation and validation}
\label{sec:impl-graph}
\label{sec:impl-valid}

Each node records an identifier, label, canvas position, and absolute local
significance level $\alpha_i$. Each directed edge records its endpoints and
transition weight. When a testing object is created, the edge table is
converted to a square transition matrix indexed by hypothesis labels.
The application queries the testing object for current allocations and
transition weights after each rejection. Graph reduction can create an edge
absent from the initial graph, so the display queries all surviving ordered
pairs, including edges created by graph reduction. Identifiers link schedule
rows to nodes; unique hypothesis labels index the backend matrix.

Node levels must be finite and nonnegative, and their total cannot exceed
one. Edge weights must be finite values in $[0,1]$, and self-loops are
excluded. At object construction, outgoing weights must sum to zero or one.
This interface constraint is stricter than the mathematical requirement
$\sum_j g_{ij}\leq 1$; a rejected node either has no successor or passes its
entire level to the remaining graph. Initial local levels are absolute
significance levels; outgoing weights are redistribution proportions.
The pinned backend rounds some recycled levels below $10^{-5}$ upward.
A guarded subclass rejects such an update before mutation, preserving the
specified allocation and the backend's behaviour above this numerical limit.

Import accepts omitted layout fields and legacy label-based endpoints.
Duplicate node identifiers, duplicate hypothesis labels, invalid alpha
allocations, duplicate directed edges, and unknown endpoints are rejected.
Schedule validation checks the number of looks, their order,
and increasing information fractions ending at one.

\Needspace{8\baselineskip}
\subsection{Boundary and testing interfaces}
\label{sec:impl-integration}

\pkg{TrialSimulator} supplies the graphical testing object and alpha
propagation \citep{TrialSimulator2026}. The application supplies a local-test adapter using \pkg{rpact}
\citep{rpact2026,WassmerBrannath2025}. Table~\ref{tab:engines}
maps the five interface choices to their spending profiles and boundary codes.
The \code{graphmtp\_local\_spec()} helper fixes the local schedule and
profile; \code{graphmtp\_local\_boundaries()} supplies both previews and
submitted cutoffs through \code{rpact::getDesignGroupSequential()}.
The integration uses \pkg{TrialSimulator}~1.35.8 and
\pkg{rpact}~4.4.0. The runtime compares the observed one-sided $p$~value
with its nominal cutoff using a strict inequality; equality has probability
zero under the continuous model in Section~\ref{sec:sequential-scope}.

\begin{table}[H]
\centering\small
\begin{tabular}{@{}P{.14\linewidth}P{.23\linewidth}P{.11\linewidth}P{.17\linewidth}P{.23\linewidth}@{}}
\toprule
UI rule & Spending profile & Boundary code & Spending schedule & Allocation update \\
\midrule
O'Brien--Fleming & \pkg{gsDesign}, \code{sfLDOF} & \code{asOF} & Locked information fraction & Reconstruct at current alpha \\
Pocock & \pkg{gsDesign}, \code{sfLDPocock} & \code{asUser} & Locked information fraction & Reconstruct at current alpha \\
Hwang--Shih--DeCani & \pkg{gsDesign}, \code{sfHSD}; shape $\eta$ & \code{asUser} & Planned look's cumulative proportion & Stored proportions times current alpha \\
HP-derived profile & Fixed interim $p_1$; normal-integration final boundary & \code{asUser} & Planned look's HP-derived proportion & Fixed initial proportions times current alpha \\
Custom & First-crossing solver; supplied cumulative amounts & \code{asUser} & Planned look's supplied proportion & Stored proportions times current alpha \\
\bottomrule
\end{tabular}
\caption{One-sided rule mapping. Previews and submitted tests use the same adapter.
All options use the complete locked information schedule to calculate
within-hypothesis correlations and reconstruct the local family at every look.}
\label{tab:engines}
\end{table}

The initial Haybittle--Peto profile follows the fixed-interim-cutoff approach
\citep{Haybittle1971}: it holds a specified nominal interim cutoff constant
and solves for a final boundary using canonical normal crossing probabilities.
The interim cutoff must be positive and smaller than the local alpha, and
the combined interim crossing probability must leave a positive final
budget. A single-look design uses the full local alpha directly.
Custom profiles are specified by a strictly increasing sequence of
positive cumulative spending values ending at the initial local alpha.
Writing the supplied amounts as $A_k$, the conversion is
$c_k=A_k/\alpha_i^{(0)}$. At any calculation with allocation $\gamma$,
the adapter uses $\gamma c_k$. Strict positivity and increase are interface
restrictions. A zero-initial-alpha hypothesis cannot specify this absolute
custom profile.
Multi-look Haybittle--Peto designs also require positive initial alpha so
that the absolute interim cutoff is admissible before the design is locked.

Hwang--Shih--DeCani spending
\citep{HSD1990} uses the planned look's proportion
$(1-\exp(-\eta t_k))/(1-\exp(-\eta))$, with limit $t_k$ at $\eta=0$.

For HP-derived spending, the initial fixed-interim-cutoff preview determines
a cumulative spending profile at the initial allocation. The application
stores the corresponding proportions when the design is finalised. At every
later allocation, it scales this same profile by the current alpha and
reconstructs the complete boundary family. Thus the profile is fixed across
looks and retests. Its nominal interim cutoffs can change after recycling;
the HP-derived label distinguishes it from a rule that keeps the interim
cutoff fixed at every allocation.

At every testing event, the adapter uses the complete prespecified
information schedule and current allocation to calculate all local
boundaries, including the final boundary. It applies the current-look cutoff
to the current $p$~value. Following rejection, it repeats this calculation for
remaining hypotheses in the submitted batch, reusing their inputs. Earlier
observed $p$~values are not re-evaluated. Earlier event rows retain the
allocation and cutoff actually used, even when the local boundary family
is subsequently reconstructed at a higher allocation.

The planned information counts are obtained by rounding each entered
fraction times the planned maximum. The effective fractions are these
counts divided by that maximum; they must increase strictly and end at one.
Submitted information counts must equal the locked planned counts. An
underrun, overrun, or other schedule change is rejected before testing.
The sequential implementation supports total family alpha at most
$0.3173105$, positive local alpha at least $10^{-5}$, and first-crossing
spending increments at least $10^{-8}$. These are numerical domain
restrictions. Failed submissions restore the preceding valid session state.

\subsection{Submitted state and persistence}
\label{sec:impl-state}
\label{sec:impl-io}

The finalised graph, initial local levels, spending plan, and analysis
schedule are serialised into a design signature. Later submissions are
checked against that signature. The signature excludes canvas positions,
which do not affect the testing procedure. Before testing a batch, the
application validates its inputs and takes a serialised deep copy of the
mutable testing object.
If testing or extraction of the event history fails, that snapshot is
restored.

The event history records the boundaries and alpha allocations used at each
submission. Later recycling updates the live state without rewriting those
stored rows. An exported JSON file contains the original graph, plan,
schedule, and event history, together with format and software version
metadata. The current format is JSON version 3 and records execution
semantics as \code{fixed-information-nested-spending-v1}. Import replays
compatible submitted inputs to reconstruct the live object. Histories from
earlier execution semantics restore the design only, with a warning, so
earlier boundaries are not silently reinterpreted.
Exports preserve stored numerical values without display rounding and retain requested cumulative
spending proportions separately from the backend's calculated spending.
Replay identifies inputs by submission, hypothesis, and the supplied
$p$~value, information, final-look flag, maximum, and spending proportion.
Repeated decisions remain distinct event rows. If replay fails, the
application clears the partial testing object and reports that only the
design was restored. Successful replay restores the design lock for both
partial and completed sessions.

\Needspace{7\baselineskip}
\section{Worked examples}
\label{sec:examples}

Four cases illustrate fixed-design graph updates and sequential monitoring.
The graphs follow the weighting strategies discussed by \citet{Bretz2011}.
All examples use an overall one-sided level of $0.025$. Their design files are
included in the replication materials and can be imported through the
\code{Design} tab. The examples use illustrative $p$~values and design inputs.

\Needspace{6\baselineskip}
\subsection{A fixed-design procedure with two primary hypotheses}
\label{sec:ex-fixed}

The first graph allocates $\alpha_1=\alpha_2=0.0125$ to two primary hypotheses
and $\alpha_3=0$ to a secondary hypothesis. Every directed edge between distinct
hypotheses has weight $0.5$. This design permits separate claims for the
primary hypotheses and makes the secondary testable after either primary is
rejected.

\begin{figure}[H]
\centering
\begin{minipage}[t]{0.32\linewidth}\centering
\small Initial allocation\\[3pt]
\includegraphics[width=\linewidth]{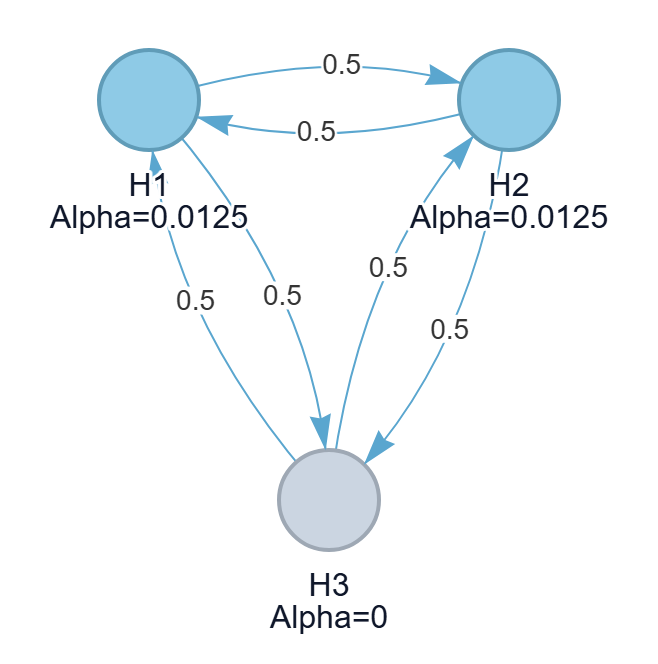}
\end{minipage}\hfill
\begin{minipage}[t]{0.32\linewidth}\centering
\small After rejecting $H_1$\\[3pt]
\includegraphics[width=\linewidth]{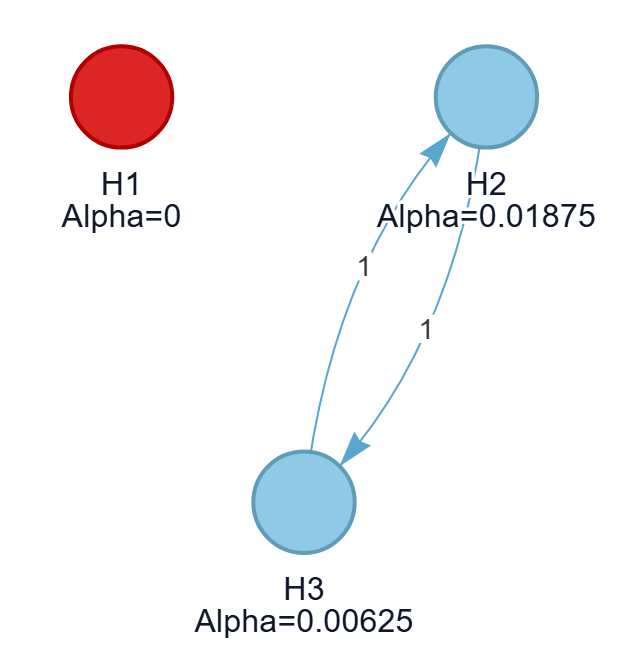}
\end{minipage}\hfill
\begin{minipage}[t]{0.32\linewidth}\centering
\small After rejecting $H_2$\\[3pt]
\includegraphics[width=\linewidth]{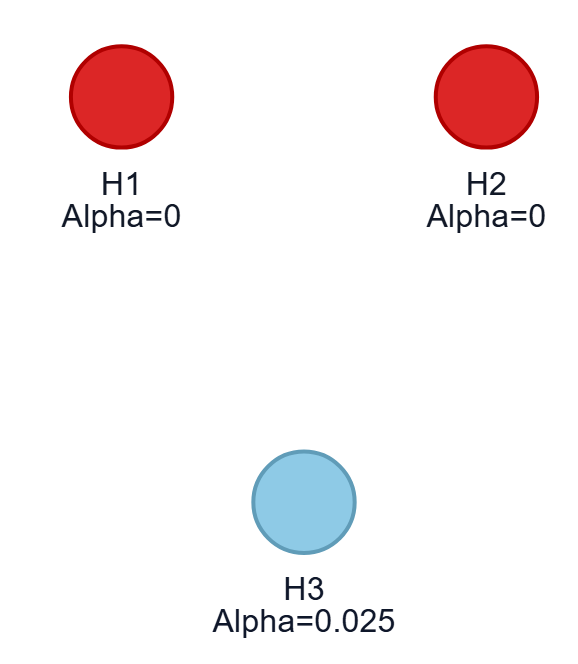}
\end{minipage}
\caption{Fixed-design graph before and after rejecting $H_1$ and $H_2$.
Red nodes denote rejections; grey denotes the initially untestable secondary.
These app-canvas views show the resulting local levels and transition weights.}
\label{fig:fixed-recycling}
\end{figure}

Suppose $H_1$ is significant at its initial local level. Rejecting it sends
half of $0.0125$ to each survivor, giving $(\alpha_2,\alpha_3)=(0.01875,0.00625)$.
The updated edges between $H_2$ and $H_3$ each have weight $1$. If $H_2$ is then
significant at $0.01875$, its rejection transfers that level to $H_3$, which
reaches $0.025$. Figure~\ref{fig:fixed-recycling} displays these states. The
graph-exploration controls follow Section~\ref{sec:workflow}.

\Needspace{10\baselineskip}
\subsection{Fixed-sequence gatekeeping}
\label{sec:ex-gatekeeping}

The second graph is the chain $H_1\to H_2\to H_3$, with transition weights
equal to $1$ and initial levels $(0.025,0,0)$.
The primary hypothesis $H_1$ must be rejected
before $H_2$ can be tested, and $H_2$ must be rejected before $H_3$ can be tested.
The successive allocations are $(0,0.025,0)$ and $(0,0,0.025)$.
The full level remains in the active graph until the terminal hypothesis
$H_3$ is rejected, when the remaining allocation becomes zero.

\Needspace{8\baselineskip}
\subsection{Group-sequential monitoring with alpha recycling}
\label{sec:ex-gs}

The third graph assigns $0.025$ to $H_1$ and zero to $H_2$, with a single
edge $H_1\to H_2$ of weight $1$. The primary has three planned looks with
Lan--DeMets O'Brien--Fleming spending. Its planned maximum information is
$393$ units, and the scheduled cumulative information values are $205$,
$285$, and $393$. Thus, the information fractions are $205/393$, $285/393$,
and $1$. The secondary has one planned look at analysis time 3. Information
is expressed in abstract units here; patient or event counts can be used
only when they are proportional to the information in the test statistic.
The example uses the same maximum in the saved design and the runtime inputs.

\begin{figure}[H]
\centering
\includegraphics[width=0.8\linewidth]{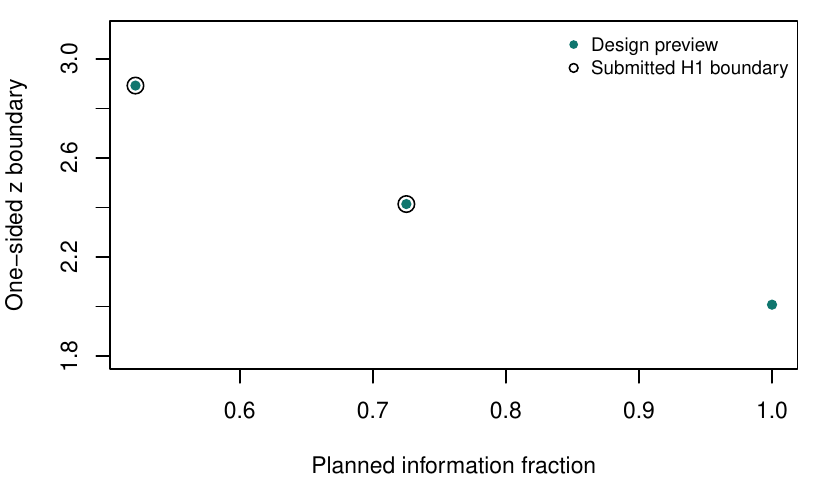}
\caption{Planned one-sided boundaries for $H_1$, calculated by \code{code.R}
with Lan--DeMets O'Brien--Fleming spending and maximum information $393$.
Open circles identify submitted tests. The third planned look is unused
because $H_1$ is rejected at look 2. Points represent discrete looks.}
\label{fig:sequential-boundaries}
\end{figure}

Figure~\ref{fig:sequential-boundaries} shows the primary's planned boundaries.
The first two submitted analyses use the planned information values,
so their runtime boundaries agree with the corresponding preview values to
the displayed precision. With $p=0.09$ at the first look, $H_1$ is not rejected
at its cutoff of $\FirstBoundaryP$. With $p=0.006$ at the second look, it is
rejected at $\SecondBoundaryP$. Its $0.025$ then transfers to $H_2$.
At analysis time 3, $H_2$ is tested once with $p=0.02$ and rejected at $0.025$.
There is no third submitted look for $H_1$ after its rejection. The last
planned point is therefore unused.

\begin{table}[H]
\centering
\small
\begin{tabular}{rlrrrrl}
\toprule
Time & Hypothesis & Information & $z$ cutoff & $p$ cutoff & Observed $p$ & Decision \\
\midrule
\SubmittedHistoryRows
\bottomrule
\end{tabular}
\caption{Submitted results extracted from the application server for the
group-sequential example; Time denotes the analysis-time index. The local level at each listed submission is
$0.025$. Cutoffs are rounded for display; decisions use unrounded values.}
\label{tab:submitted-results}
\end{table}

Table~\ref{tab:submitted-results} reports testing events recorded by the
application server. The replication script
also replays the saved session and compares the resulting trajectory with
the original. It independently applies
\code{graphicalMCP::graph\_test\_shortcut\_gsd()} with the same graph,
information fractions, ordinary one-sided $p$~values, and
\code{look\_back = FALSE}. Both implementations reject $H_1$ at analysis
time 2 and $H_2$ at analysis time 3. Agreement was assessed for the rejection
sequence of this graph and schedule.

\subsection{Recycling to a previously monitored hypothesis}
\label{sec:ex-recipient}

A second sequential case follows two monitored treatment comparisons.
Both $H_1$ and $H_2$ initially receive $0.0125$, have maximum information
$100$, and use O'Brien--Fleming spending at fractions $0.33$, $0.67$, and $1$.
The graph transfers the full level of a rejected $H_2$ to $H_1$.
This represents two initially testable claims with a prespecified direction
of recycling. The case extends the production-server regression fixture
to include a recipient with an earlier nonrejection.

At analysis time 1, both hypotheses have $p=0.5$ and remain active.
At time 2, the application first tests $H_1$ with information $67$ and
$p=0.008$. It is not rejected at cutoff $\RecipientOldCutoff$.
The rejection of $H_2$ with $p=0.0001$ then increases $H_1$'s allocation
from $0.0125$ to $0.025$. Retesting the unchanged $H_1$ input against
$\RecipientNewCutoff$ also does not reject it. At time 3, $H_1$ has information
$100$ and $p=0.02$, which crosses its final cutoff of $0.0230919$
(Table~\ref{tab:recipient}).
The first-look row remains at its original allocation and cutoff.
The retest reconstructs the interim boundary calculation at the higher
allocation, including the information times of previous looks.

\begin{table}[H]
\centering\small
\setlength{\tabcolsep}{4pt}
\begin{tabular}{rrlrrrrl}
\toprule
Event & Time & Hypothesis & Local alpha & Info. & Observed $p$ & $p$ cutoff & Decision \\
\midrule
\RecipientRows
\bottomrule
\end{tabular}
\caption{Recorded events for a previously monitored recipient of alpha.
Events 3 and 5 use the same second-look data for $H_1$; event 4 triggers
the allocation change. Event 6 tests $H_1$ at its final look. ``Retain''
means do not reject at that event.}
\label{tab:recipient}
\end{table}

The saved JSON reproduces all six event rows, including both second-look
comparisons and the final test of $H_1$. Independent normal integration
checks the complete reconstructed three-look family at alpha $0.025$;
its total crossing probability agrees with $0.025$ within $10^{-5}$.
This calculation evaluates the fixed local family at that allocation,
not the probability of the single illustrated sequence of observations.

\section{Validation and reproducibility}
\label{sec:validation}

The verification suite checks numerical calculations and application state
on the recorded platform. Section~\ref{sec:sequential-scope} states the
methodological conditions and the supported scope.

\subsection{Numerical checks}

The verification scripts compare the standard boundary calculations with
direct \pkg{gsDesign} calls, including several Hwang--Shih--DeCani shape
parameters. These checks assess consistency with the preview design routines;
they are not independent checks of code paths that themselves call
\pkg{gsDesign}. The verification
suite also checks the single-look limit and
evaluates the total crossing probability of a three-look Haybittle--Peto
boundary using \pkg{mvtnorm}'s deterministic Miwa integration. The latter
calculation is independent of the boundary adapter and agrees with the target $0.025$ within $10^{-5}$.
Three-look production sessions at fractions $0.33$, $0.67$, and $1$ and
unchanged local alpha $0.025$ exercise all five spending choices
and check their canonical-normal crossing probabilities at the final look
against $0.025$, with the same tolerance.

The script \code{independent\_boundary\_review.R} in the
\code{replication} directory checks the released adapter with Miwa integration
for all five spending families at allocations $0.0125$, $0.025$, and $0.05$.
It compares cumulative first-crossing probabilities with the specified
spending targets at each of three looks, using tolerance $5\times10^{-8}$.
These 15 probes provide an independent numerical check of the returned
boundaries for the tested configurations, not a uniform accuracy guarantee.

For the sequential example, \code{code.R} evaluates the canonical-normal
probability of crossing either of the first two submitted boundaries and
checks it against the Lan--DeMets cumulative spending target. Both are
$0.0084872$ to seven decimal places; the assertion tolerance is $10^{-5}$.
The fixed-design examples are also checked against an independent
implementation of the published graph-update equations, with tolerance
$10^{-12}$. The independent \pkg{graphicalMCP} calculation described in
Section~\ref{sec:ex-gs} checks the rejection sequence. The monitored-recipient
case checks reconstruction at both allocations and the final look; the focused HP
probe checks the change in cutoff within one submission. These are bounded
tests of the stated configurations.

Explicit closure tests use one acyclic graph with three hypotheses and
sixteen data configurations, plus two graphs with terminal cycles and
singular complements.
After each submitted look, runtime rejections are contained in the
full-history closed rejection set. A separate fixture demonstrates the
strict conservatism possible when only current-look data are retested.

\subsection{Application-state checks}

Production-server checks use \code{shiny::testServer()} to exercise
malformed imports, graph edges created by rejection,
design-signature changes, invalid schedules, submitted information,
same-analysis recycling, delayed activation, and saved-session replay.
Failure-injection tests check that a backend error cannot leave a partially
mutated testing object. Upload-path tests check that successful replay
restores the design lock for both partial and completed sessions, while
design-only imports and inconsistent histories remain unlocked.
Replay compares recalculated results with the
saved local levels, cumulative spending, $z$ and $p$ cutoffs, and observed
$p$~values with absolute tolerance $10^{-6}$. Hypothesis identifiers and
decisions must agree exactly. An inconsistent history is rejected.
Numerical-domain guards are tested for failure before state mutation.

Lightweight helper fixtures complement the production-server cases.
The server-session helper submits inputs to the production collection and
history functions; separate upload-observer tests cover import and locking.
Browser capture supplies evidence for the displayed interface states.
The validation directory records each category and its actual transcript.

\subsection{Replication materials}

The accompanying \code{code.R} is the single entry point for the reported
numerical results and the reference boundary plot. It also produces vector
reference diagrams for checking the two fixed-design graphs. It reads the shipped design
files, runs the sequential example through the application server, writes
the table fragment included by this manuscript, and saves a completed
JSON session. From the top-level replication bundle directory, it can be run in
\proglang{R} as follows.

\begin{CodeChunk}
\begin{CodeInput}
R> source("code.R")
\end{CodeInput}
\end{CodeChunk}

The companion \code{code.html} shows code and results together.
The \code{capture\_interface.R} script in the \code{replication} directory
captures the local app through file import and normal controls, with full
methods in \code{validation/FIGURE\_PROVENANCE.md}. From the same bundle root,
run \code{Rscript software/scripts/run\_verify\_all.R} for the verification
suite. Package versions, \code{sessionInfo()}, and SHA256 file hashes
identify the tested software and replication materials.

A clean dependency restore installed 106 dependencies into an initially
empty library with package caching disabled; four recommended packages
were supplied with \proglang{R}. All 110 lockfile versions matched the
installed environment. The restore used
\proglang{R}~4.5.3 on Windows 11 with Rtools45.
The restoration transcript and package-version audit accompany the
replication materials. This verifies installation of the locked environment;
the numerical and application checks above assess the reported calculations
and software behaviour.

\section{Comparison with existing tools}
\label{sec:comparison}

\pkg{graphMTP} combines an interactive graph editor, a per-hypothesis
monitoring schedule, and a record of submitted analyses. \pkg{gMCP} provides
graphical-testing methods and a desktop editor \citep{gMCP2024}.
\pkg{gMCPShiny} provides a browser interface for graphical multiplicity
control and documents hosted instances \citep{gMCPShiny2023}.

Related functionality is available in commercial trial-design environments.
\proglang{East Horizon} documents multi-arm, multi-stage designs and testing
of multiple endpoints \citep{Cytel2026}. \proglang{nQuery} documents
multiple-comparison sample-size calculations and group-sequential design
tools \citep{Statsols2026}. These capabilities are described from product
documentation; licensed installations were not empirically evaluated.

\pkg{graphicalMCP}~0.3.0 includes graphical group-sequential testing with
per-hypothesis spending functions \citep{graphicalMCP2026}. Its
\code{graph\_test\_shortcut\_gsd()} function accepts matrices of ordinary
$p$~values and information fractions, including missing entries for looks
not taken. It distinguishes current-look testing from looking back across
earlier analyses. These programmatic choices support scripted analyses and
method comparisons. \pkg{graphMTP} guides a user through individual
submissions and displays the resulting graph and testing history.

\pkg{gsDesign} focuses on group-sequential design and monitoring, while
\pkg{TrialSimulator} supplies the graphical-testing runtime used here
\citep{gsDesign2026,TrialSimulator2026}. Table~\ref{tab:comparison} summarises
documented workflow features. The comparison refers to \pkg{gMCP}~0.8-17,
\pkg{gMCPShiny}~2023.11.0, \pkg{graphicalMCP}~0.3.0,
\pkg{gsDesign}~3.11.0, and \pkg{TrialSimulator}~1.35.8, as checked in
September 2026. The installed \pkg{graphicalMCP} API supplies the
version-specific comparison; its moving website can describe later
development versions.

\begin{table}[H]
\centering\small
\begin{tabular}{@{}P{.18\linewidth}P{.15\linewidth}P{.18\linewidth}P{.21\linewidth}P{.16\linewidth}@{}}
\toprule
Software & Graph editing & Hypothesis-specific schedules & Recorded testing results & Session replay \\
\midrule
\pkg{gMCP} & Desktop and code & Not assessed & Test results & Not assessed \\
\pkg{gMCPShiny} & Browser & Not assessed & Multiplicity results & Not assessed \\
\pkg{graphicalMCP} & Code & Yes & Rejections and test values & Not assessed \\
\pkg{gsDesign} & Not assessed & Local sequential design & Boundary calculations & Not assessed \\
\pkg{TrialSimulator} & Code & Yes & Testing trajectory & Not assessed \\
\pkg{graphMTP} & Browser & Yes & Submitted-event history & JSON input replay \\
\bottomrule
\end{tabular}
\caption{Workflow comparison from the cited package materials and supplied
implementation. ``Not assessed'' indicates an unverified capability, not
absence. Recorded results have the package-specific forms listed here.}
\label{tab:comparison}
\end{table}

The integrated workflow lets collaborators inspect a graph, link each
hypothesis to its monitoring schedule, and recover the boundary used for a
particular submitted test. The worked cases demonstrate these review tasks
with executable inputs and saved histories.

\section{Discussion}
\label{sec:discussion}

\pkg{graphMTP} provides a visual workflow for specifying and inspecting
graphical multiple-testing procedures. The graph and history give
complementary views: the graph shows the current allocation, while the
history records the boundary and decision for each testing event. Saving
the design with that history supports review and replication when the
software environment is retained.

The application takes endpoint-specific one-sided $p$~values and information
counts as inputs. It does not estimate treatment effects, determine sample
size, or generate patient-level data. Its sequential scope follows
Section~\ref{sec:sequential-scope}. The monitored-recipient case illustrates
why a review of earlier decisions requires the event history as well as the
current graph.

The current interface excludes futility stopping, conditional-error
redesign, and arbitrary sample-size adaptation. The HP-derived runtime
profile requires the allocation and timing interpretation in
Section~\ref{sec:impl-integration}. These limits matter when translating a
trial protocol into an executable design. The supplied cases provide a
bounded basis for reviewing that translation and its recorded calculations.

JSON exports retain the design and analysis history; access control and
tamper detection are outside the implementation. Extensions could provide formatted
analysis reports and separately justified adaptive monitoring procedures.

\Needspace{7\baselineskip}
\section*{Computational details}
\label{sec:compdetails}

The results were generated with \proglang{R}~4.5.3 \citep{Rcore2026} and
\pkg{graphMTP}~0.3.0 on 64-bit Windows 11
(\code{x86\_64-w64-mingw32}). Statistical dependencies were
\pkg{TrialSimulator}~1.35.8, \pkg{rpact}~4.4.0, \pkg{gsDesign}~3.11.0,
and \pkg{mvtnorm}~1.4-2. The independent comparison used
\pkg{graphicalMCP}~0.3.0. The full environment is recorded in
\code{validation/sessionInfo.txt}. The replication script sets seed
\code{20260908} for numerical routines using stochastic integration;
graph arithmetic and Miwa integration are deterministic.

The accompanying software includes an MIT-licensed installable
\proglang{R} source package and the maintained standalone application.
Installation instructions accompany the environment-restoration script in
the replication directory. The package provides a documented launcher,
\code{run\_app}. From the bundle root, the standalone application
starts with the following command.
\begin{Code}
shiny::runApp("software")
\end{Code}
The development installer obtains current CRAN dependencies; replication
uses the versions in \code{renv.lock}.
The reference platform and numerical tolerances define the reported
calculation, without a claim of bit-for-bit cross-platform agreement.
\maintainername{} and MengYang Yi maintain the software; \contactname{} is
the package's formal contact.

A SHA256 file manifest identifies the accompanying sources and results.
The software snapshot uses commit \code{e9bd18690d53}.
Development sources are at
\url{https://github.com/valle1306/graphical-testing-procedure}.
The hosted app at
\url{https://oncotrialdesign.shinyapps.io/graphMTP/} will be updated
periodically to remain consistent with the GitHub version.

\bibliography{references}
\end{document}